\documentclass[a4paper]{jpconf}
\usepackage{graphicx}

\usepackage{latexsym}
\usepackage{amsmath,,calrsfs}
\usepackage{amsfonts}
\usepackage{amssymb}
\usepackage{amscd}
\usepackage{fancybox}
\usepackage{cite}
\usepackage{amsmath,amsfonts,amsbsy}
\usepackage{pstricks,pst-node}
\usepackage[small,bf,hang]{caption2}
\usepackage{graphicx}
\usepackage{epsfig}
\usepackage{psfrag}

\usepackage{float}

\psset{unit=1.3cm,linewidth=.5pt,radius=.2}  

\usepackage{multirow}                     
\usepackage{float}                          
\usepackage{lscape}                         
\usepackage{bm}

\newcommand{\cD}{{\cal D}}

\newcommand{\beq}{\begin{equation}}
\newcommand{\eeq}{\end{equation}}
\newcommand{\ei}{\end{itemize}}
\newcommand{\bt}{\begin{tabular}}
\newcommand{\et}{\end{tabular}}
\newcommand{\bc}{\begin{center}}
\newcommand{\ec}{\end{center}}

\newcommand{\ft}[2]{{\textstyle {\frac{#1}{#2}} }}

\newcommand{\be}{\begin{equation}}
\newcommand{\ee}{\end{equation}}
\newcommand{\bea}{\begin{eqnarray}}
\newcommand{\eea}{\end{eqnarray}}
\newcommand{\ba}{\begin{array}}
\newcommand{\ea}{\end{array}}

\def\bbox{{\,\lower0.9pt\vbox{\hrule \hbox{\vrule height 0.2 cm
\hskip 0.2 cm \vrule height 0.2 cm}\hrule}\,}}
\newcommand{\dsl}{\pa \kern-0.5em /}

\makeatletter \@addtoreset{equation}{section} \makeatother

\def\slashchar#1{\setbox0=\hbox{$#1$}           
   \dimen0=\wd0                                 
   \setbox1=\hbox{/} \dimen1=\wd1               
   \ifdim\dimen0>\dimen1                        
      \rlap{\hbox to \dimen0{\hfil/\hfil}}      
      #1                                        
   \else                                        
      \rlap{\hbox to \dimen1{\hfil$#1$\hfil}}   
      /                                         
   \fi}

\def\eq#1{(\ref{#1})}

\def\tH{{\widetilde H}}

\begin{document}
\rightline{ UG-10-35 , MIT-CTP-4187 , DAMTP-2010-81, MIFP-10-44}
\title{On Critical Massive (Super)Gravity in adS$_3$}

\author{Eric A.~Bergshoeff ${}^1$, Olaf Hohm ${}^2$, Jan Rosseel ${}^1$, Ergin Sezgin ${}^3$ and Paul K.~Townsend ${}^4$ }

\address{${}^1$ Centre for Theoretical Physics, University of Groningen,  Nijenborgh 4, 9747 AG Groningen, The Netherlands}

\address{${}^2$ Center for Theoretical Physics, Massachusetts Institute of Technology,  Cambridge, MA 02139, USA}

\address{${}^3$ George and Cynthia Woods Mitchell Institute for Fundamental Physics and Astronomy, Texas A\&M University, College Station,
TX 77843, USA}

\address{${}^4$ Department of Applied Mathematics and Theoretical Physics, Centre for Mathematical Sciences, University of Cambridge,
Wilberforce Road, Cambridge, CB3 0WA, U.K.}

\ead{E.A.Bergshoeff@rug.nl, ohohm@mit.edu, j.rosseel@rug.nl, sezgin@tamu.edu, P.K.Townsend@damtp.cam.ac.uk}

\begin{abstract}

We review the status of  three-dimensional  ``general massive gravity'' (GMG)  in its linearization about an anti-de Sitter (adS) vacuum, focusing on critical points in parameter space that yield generalizations of  ``chiral gravity''. We then show how these results extend to $\mathcal{N}=1$ super-GMG, expanded about a supersymmetric adS vacuum, and also to the most general
`curvature-squared' $\mathcal{N}=1$ supergravity model.

\end{abstract}


Massive gravity models in three spacetime dimensions (3D) have been intensively investigated over the past few years  because of the possibility of finding a consistent `toy' model of quantum gravity from which we might learn something useful. The oldest and simplest of these massive gravity models is ``topologically massive gravity'' (TMG), which is defined by the addition  to the usual Einstein-Hilbert (EH) action of a Lorentz Chern-Simons (LCS) term  \cite{Deser:1981wh}. TMG propagates a single massive spin-$2$ mode but this is a ghost unless the EH term has the non-standard sign. It is useful  to define a sign $\sigma$ such that $\sigma=1$ yields the standard EH term, so that  ``non-standard'' means $\sigma=-1$. The addition of a cosmological term allows the possibility of anti-de Sitter (adS) vacua,  but  then $\sigma=-1$ implies a negative mass for  (BTZ) black holes. In the context of the adS$_3$/CFT$_2$ correspondence, this problem with the bulk theory translates to a negative central charge of the  boundary conformal field theory (CFT). Taking $\sigma=1$ allows positive central charges but at the cost of non-unitary propagation of the bulk spin-$2$ modes, which again implies \cite{Skenderis:2009nt,Grumiller:2009mw}, although less directly,  non-unitarity of the  boundary CFT.

The classical equations of TMG depend on a mass parameter $\mu$ associated to the LCS term, and a length scale $\ell$ associated to the cosmological term,  which we define such that $\ell$ is the radius of curvature in an adS vacuum.  However, rescaling the metric is equivalent to rescaling $\mu$ and $\ell$ so the only parameter of the classical theory is the dimensionless product $\mu\ell$. Quantum corrections will introduce the additional dimensionless constant $\kappa^2/\ell$, where  $\kappa= \sqrt{16\pi G_3}$ is the gravitational coupling constant, but this appears classically only through an overall  factor of  $\ell/\kappa^2$ in the on-shell action.  In the equivalent approximation to the boundary CFT,  the central charges take the form $c_\pm = (24\pi\ell/\kappa^2)f_\pm(\mu\ell)$
for dimensionless functions $f_\pm$.   It was pointed out in \cite{Li:2008dq} that the parameter $\mu\ell$  can be tuned to a critical value, the ``chiral point'', at which either $c_+$ or $c_-$ is zero. The bulk massive graviton mode then disappears from the spectrum, which  suggests  that the problem of a non-unitary bulk graviton for $\sigma=1$  is circumvented by this ``chiral gravity'' theory. The definition of such a theory depends crucially on a choice of  boundary conditions \cite{Grumiller:2008qz,Maloney:2009ck}.  Boundary conditions weaker than the standard Brown-Henneaux boundary conditions  lead to well-defined logarithmic boundary CFTs. They also allow new logarithmic bulk modes, which is  related to the observation of \cite{Carlip:2008jk} that  the bulk spin-$2$ mode  is  replaced by a bulk spin-$1$ mode at the chiral point.

An alternative, parity-preserving 3D massive gravity is ``new massive gravity'' (NMG),  which is defined by the
addition to the EH plus cosmological terms of a curvature-squared term constructed from the scalar \cite{Bergshoeff:2009hq}
\be\label{curvsq}
G^{\mu\nu} S_{\mu\nu} \equiv R^{\mu\nu}R_{\mu\nu} - \frac38 R^2
\ee
where the tensors $G,S,R$ are, respectively, the Einstein, Schouten and Ricci tensors. This  involves the introduction of a mass-squared  parameter $m^2$ (which is positive for NMG but which we allow to be negative in the most general curvature-squared model). In an expansion about a Minkowski vacuum one finds that two spin-$2$ modes of mass $m$ are propagated and, as for TMG, that perturbative unitarity  requires $\sigma=-1$. As required by parity, the two spin-$2$ modes have opposite sign ``helicities'' $s$; i.e. opposite sign of the 3D Pauli-Lubanski pseudo-scalar divided by the mass; we call $|s|$ the ``spin''.

In the adS context, there is a family of classical NMG models parametrized by the dimensionless constant $m^2\ell^2$.  In any parity-preserving 3D gravity theory with an adS vacuum, the central charges of the  boundary CFT are equal, $c_+=c_-=c$, and may be computed by the formula \cite{Brown:1986nw,Saida:1999ec,Kraus:2005vz}
\be\label{parityeven}
c= \left. \frac{8\pi\ell}{\kappa^2} g_{\mu\nu}\frac{\partial L_{3D}}{\partial R_{\mu\nu}} \right|_{\rm adS \ vacuum}
\ee
where $L_{3D}$ is the 3D gravity Lagrangian of mass dimension two, i.e. without the factor of $\kappa^{-2}$. For NMG this  gives
\cite{Liu:2009bk,Bergshoeff:2009aq,Hohm:2010jc}
\be
c = \frac{24\pi\ell}{\kappa^2}\ \hat\sigma\, , \qquad \hat\sigma = \sigma + \frac{1}{2\ell^2 m^2}\, .
\ee
The expansion of the NMG action about the adS vacuum is greatly facilitated by starting with an alternative second-order version of the action with an  auxiliary tensor field  \cite{Bergshoeff:2009hq}.  Diagonalization of the quadratic term of this action yields the sum  of a linearized EH action with coefficient $\hat\sigma$,  which is therefore the `effective' EH coefficient in the adS background, and a spin-$2$ Fierz-Pauli (FP) action with coefficient $-1/\hat\sigma$ \cite{Bergshoeff:2009aq}. The linearized EH action propagates no modes so we get two massive gravitons from the FP action, but  these  are propagated unitarily only if $\hat\sigma<0$. Hence there is a clash between unitary bulk gravitons and positive CFT central charges, as in TMG. There is also an analog of chiral gravity because there is a critical value of $\ell^2 m^2$ at which $c=0$, and at which the quadratic action becomes equivalent to a spin-$1$ Proca action \cite{Bergshoeff:2009aq}.  This critical NMG model is associated to a logarithmic CFT \cite{Grumiller:2009sn}.

Both TMG and NMG are special cases of ``general massive gravity''
(GMG) \cite{Bergshoeff:2009hq},  defined by the addition  of {\it
both} the LCS and  the NMG curvature-squared term to the EH plus
cosmological terms.  Expanded about a Minkowski  vacuum, this model
propagates two spin-$2$ modes of opposite helicity but with
different masses.  In an adS vacuum there is a two-parameter family
of classical GMG models parametrized by the constants  $\mu\ell$ and
$m^2\ell^2$. The central charges of the boundary CFT can be computed
by a generalization of (\ref{parityeven}) to allow for parity-odd
terms  \cite{Kraus:2005zm}. The result is
\be\label{ccfun} c_\pm =
(24\pi\ell/\kappa^2) f_\pm \, , \qquad f_\pm= \hat\sigma \pm
\frac{1}{\mu\ell} \, . \ee
The no-ghost conditions follow from an
expansion of the GMG action about an adS vacuum, but this
`off-shell' analysis is more subtle here, in part because the
auxiliary tensor field `trick' reduces the order in derivatives to
three rather than two when the LCS term is present. Recent results
in \cite{Grumiller:2010tj} indicate that the parameter ranges with
positive central charge and positive energy of the graviton modes
are mutually exclusive in GMG, as in TMG and NMG. Here we shall
review, with some simplifications, the results of the on-shell
analysis of \cite{Liu:2009pha} and explain how it extends to
supergravity.

To facilitate our task,  it will be convenient to first review the
connection between a unitary irreducible representation (UIR) of the
$SO(2,2)$ isometry group of adS$_3$ and fields on adS$_3$.  Let
$\varphi^{(s)}$ be a rank-$|s|$ ($|s|\ge1$) totally-symmetric and
tracefree field on adS$_3$ subject to the `divergence-free'
condition \be \bar D^\mu \varphi^{(s)}_{\mu\nu_1\cdots \nu_{s-1}}
=0\, , \ee where $\bar D$ is the covariant derivative with respect
to a background adS$_3$ metric $\bar g$. Let  $\cD(\eta)$ be the
family of first-order linear differential operators, parametrized by
a dimensionless constant $\eta$, that act on the space of such
tensors according to the definition \be\label{hs} \left[ {\cal
D}(\eta) \varphi^{(s)}\right]_{\mu_1\cdots \mu_s} = \left[{\cal
D}(\eta)\right]_{\mu_1}{}^\rho \varphi^{(s)}_{\rho\mu_2 \cdots
\mu_s}\, , \qquad \left[
\mathcal{D}\left(\eta\right)\right]_\mu{}^\nu = \ell^{-1}
\delta_\mu^\nu + \frac{\eta}{\sqrt{|\bar
g|}}\varepsilon_\mu{}^{\tau\nu}\bar D_\tau\, . \ee Despite
appearances, the rank-$|s|$ tensor $\cD(\eta)\phi^{(s)}$  is also
traceless, totally symmetric  and `divergence-free'. Using the
identity \be\label{ident} \cD(\eta)\cD(-\eta)\varphi^{(s)} \equiv -
\eta^2 \left[\bar D^2 + \ell^{-2}\left(|s|
+1-\eta^{-2}\right)\right]\varphi^{(s)}\, \ee we see that
eigenfunctions of  the  covariant D'Alembertian $\bar D^2$ acting on
spin-$|s|$ fields are linear combinations of  solutions to the
first-order equations \be\label{basic2} \cD(\eta)\varphi^{(s)} =0 \,
, \qquad \cD(-\eta)\varphi^{(s)} =0 \, . \ee A physically
acceptable\footnote{i.e. nonsingular at the origin and normalizable
with respect to the $SO(2,2)$ invariant measure
\cite{RLN,Breitenlohner:1982jf}.} solution of either equation
furnishes a UIR of $SO(2,2)$ labelled by its lowest weights
$(E_0,s)$, where $\ell^{-1}E_0$ is the lowest energy\footnote{Or
highest energy. It is not possible to distinguish between the two at
the level of the equations of motion. This is why the unitarity of
the irreducible representation is not sufficient for unitarity of
the field theory.}, which is real and satisfies the condition $E_0
\ge |s|$ \cite{RLN,Breitenlohner:1982jf,Barut:1986dd}.  The UIR
furnished by a solution of (\ref{basic2})  has lowest weights
\cite{Deger:1998nm}
\be
E_{0} = 1+\frac{1}{|\eta|}\ ,\qquad s= \frac{|s|\eta}{|\eta|}\ .
\label{Eeta}
\ee
We see that the sign of $\eta$ gives the sign of the helicity $s$.
The UIR's with $E_0=|s|$ for $s\ne0$ are called singleton irreps; their weight space is drastically reduced compared to a typical UIR, and they can be interpreted as describing modes that are confined to the 2D boundary of adS$_3$ \cite{HarunarRashid:1991bv}. More generally, the adS/CFT correspondence assigns to every UIR an operator in a dual 2D CFT with conformal weights  $h_\pm =(E_0\pm s)/2$, in which context the bound $E_0 \ge |s|$ translates into $h_\pm \ge 0$.

The above analysis applies for $|s|>0$. For $s=0$ we must consider the solutions of the second-order equation
\be\label{KGeq}
\left(\bar D^2 - \mathcal{M}^2\right)\varphi =0
\ee
for constant $\mathcal{M}$. The solutions, if physically acceptable,  furnish UIRs  with $s=0$ and
\be\label{spinzero}
E_0 = 1 \pm \sqrt{1+ \ell^2\mathcal{M}^2}\, .
\ee
Reality of $E_0$ implies the Breitenlohner-Freedman (BF) bound $\ell^2\mathcal{M}^2 \ge-1$.  In addition, $E_0\ge0$ is required for a UIR. This condition allows both signs in (\ref{spinzero}) when $-1\le \ell^2\mathcal{M}^2\le 0$; otherwise only the plus sign is allowed.

Let us now see how all this applies to the  `Einstein' theory
defined by the EH and cosmological terms alone. We  write the metric
as \be\label{perturb} g_{\mu\nu} = \bar g_{\mu\nu} \left(1+
\frac{1}{3}h\right) +  H_{\mu\nu}\, , \qquad \bar g^{\mu\nu}
H_{\mu\nu} = 0\, , \ee so that $(H,h)$ are the tracefree and trace
perturbations respectively.  In the gauge
\be\label{gaugecon} \bar D^\mu H_{\mu\nu} =0\, , \ee the linearized
field equations may now be written as \be\label{Einstein}
\left[\cD(1)\cD(-1)H\right]_{\mu\nu} =  -\frac{1}{3}  \left(\bar
D_\mu \bar D_\nu - \frac{1}{3} \bar g_{\mu\nu} \bar D^2\right) h\, ,
\qquad  \left(\ell^2\bar D^2 -3\right)h = 0\, . \ee We have coupled
equations for $H$ and $h$ but there is still a  residual gauge
invariance. If we write the vector gauge parameter of linearized
diffeomorphisms as $\xi^T_\mu + \partial_\mu \xi$, where $\bar D^\mu
\xi_\mu^T =0$, then the gauge condition (\ref{gaugecon}) is
invariant provided that \be \bar D_\mu \left(\ell^2\bar D^2
-3\right)\xi =0 \, , \qquad \left(\ell^2\bar D^2-2\right) \xi_\mu^T
=0 \quad \Rightarrow \quad \left(\ell^2 \bar D^2 +2\right) \bar
D_{(\mu} \xi_{\nu)}^T =0\, . \ee The variation of $h$ is $\delta h
\sim \bar D^2 \xi$ but this implies $\delta h \sim \xi$ for ${\bar
D}^2 \xi \sim \xi$. Since $\xi$  is a  residual gauge parameter, we
may use it to set $h$ to zero.  This leaves us with the equation
$\cD(1)\cD(-1)H=0$,  but the identity (\ref{ident}) implies that
this is equivalent to \be \left(\ell^2\bar D^2 +2\right) H=0\, . \ee
This is clearly invariant under the residual gauge invariance, which
suffices to  remove all local degrees of freedom of $H$.

We are now in a position to discuss GMG.  Linearized about an adS vacuum, the GMG equations are
\begin{eqnarray}\label{Einstein2}
\left[\cD(1)\cD(-1)\cD(\eta_+) \cD(\eta_-) H\right]_{\mu\nu} &=&  -\frac{1}{3\ell^2} \left(\bar D_\mu \bar D_\nu - \frac{1}{3} \bar g_{\mu\nu} \bar D^2\right) h\, , \\
\frac{\Omega}{m^2}  \left(\ell^2\bar D^2 -3\right)h &=& 0\, ,
\end{eqnarray}
where \be\label{Omega} \Omega \equiv \ell^2 m^2 \hat\sigma -1 \ee
and \cite{Liu:2009pha} \be \eta_\pm = \Omega^{-1} \left(-\frac{\ell
m^2}{2\mu}\pm \sqrt{\frac{\ell^2 m^4}{4\mu^2}-\Omega}\right)\ .
\label{etapm1} \ee We have retained the factor of $\Omega$ in the
$h$-equation because it allows the $\Omega=0$ equations to be
obtained by taking the $\Omega\to0$ limit. The region in parameter
space with $\Omega>0$ is divided from the region with $\Omega<0$ by
the curve $\Omega=0$. To see the significance of this division,  we
observe that \be\label{etaprod} \eta_+ \eta_- = \Omega^{-1}\, . \ee
This shows that the parameters $\eta_\pm$ have opposite signs in the
$\Omega<0$ region  and the same sign in the $\Omega>0$ region. The
reality of $\eta_\pm$ and the UIR condition $|\eta_\pm|\le 1$
translate into $m^2 c_\pm \le 0$ for $\Omega<0$ and  $m^2 c_\pm \ge
0$, $\ell^2 m^4 \ge 4\mu^2\Omega$ for $\Omega>0$.

Provided that $\Omega\ne0$, we may proceed as in the `Einstein'
case, eliminating $h$ by a combination of its field equation and a
residual gauge transformation, to arrive at the single  4th order
equation \be\label{keyGMG} \cD(1)\cD(-1)\cD(\eta_+) \cD(\eta_-) H=
0\, . \ee As long as $\left(\eta_+-\eta_-\right)\left(|\eta_+|
-1\right)\left(|\eta_-|-1\right) \ne0$ the general solution is a
linear combination of the two singleton modes and solutions of the
first-order equations $\cD(\eta_\pm) H=0$. This state of affairs
applies in the NMG limit $|\mu|=\infty$ since we then have \be
\eta_+ = -\eta_- = -1/\sqrt{-\Omega}\, . \ee Clearly, we must be on
the $\Omega<0$ side of the parameter space divide to take this
limit, in which the bound $|\eta_\pm|\le1$ becomes
$m^2\hat\sigma<0$. As confirmed by the computation of the quadratic
action in \cite{Bergshoeff:2009aq}, this bound is the condition for
the absence of tachyons. This computation also shows that
$\hat\sigma<0$ is the no-ghost condition, i.e. for unitarity of the
quantum field theory. As $m^2>0$ for NMG, the absence of tachyons
implies the absence of ghosts. It seems likely to us that this will
remain true for GMG as long as $\Omega<0$.  However, it  also seems
likely that GMG has ghosts when  $\Omega>0$, even though it is
certainly  possible  to satisfy the UIR bound $|\eta_\pm|<1$, and
hence avoid tachyons.  This is because solutions of the equation
$\cD(\eta)\cD(\eta')H=0$ with $\eta\eta'>0$ could be expected to
arise naturally in the context of 5th order 3D gravity models, which
necessarily propagate ghosts in a Minkowski vacuum
\cite{Bergshoeff:2009tb}. To settle this issue we would need an
`off-shell' analysis like that presented for NMG in
\cite{Bergshoeff:2009aq}.

When $\Omega=0$ the  $h$-equation drops out. This leaves us with the
$H$-equation, that can be shown to reduce to
\be \label{partmass} \left[\cD(\eta) {\cal D}(1){\cal
D}(-1)\right]_\mu{}^\rho \varepsilon_\rho{}^{\alpha \beta}
\bar{D}_\alpha H_{\beta \nu} = 0\, , \qquad \eta= - \frac{\mu}{\ell
m^2}\, .\ee
Unless $|\eta|=1$, the solutions space is spanned by the singletons,
the solution of $\cD(\eta)H=0$,  and the solution of \be\label{pm}
\varepsilon_\rho{}^{\alpha \beta} \bar{D}_\alpha H_{\beta \nu}=0\, .
\ee By acting on this equation with the operator
$\varepsilon_\lambda{}^{\tau\mu}\bar D_\tau$, we see that the
solution of (\ref{pm}) also obeys $(\ell^2 \bar{D}^2 + 3)H=0$.
Comparing this with (\ref{ident}), we might naively conclude that it
requires $\eta=\infty$ and hence $E_0=1$, which violates the
spin-$2$ bound $E_0\ge2$.  The resolution of this puzzle is subtle
because of an additional gauge invariance;  the relevant solutions
of \eq{pm} are ``partially massless''
\cite{Deser:1983mm,Deser:2001us}.

Our main interest here is  in the `critical' GMG cases. For finite
non-zero $\Omega$ these are those for which \be
\left(\eta_+-\eta_-\right)\left(|\eta_+|
-1\right)\left(|\eta_-|-1\right) =0\, . \ee When this  condition
holds, the solution space is no longer spanned by solutions of
first-order equations of the form $\cD(\eta)H=0$. However, subject
to appropriate boundary conditions, one finds that additional
solutions appear. In the case that two of the $\eta$ values
coincide, these  are logarithmic solutions of  $\cD^2(\eta)H=0$ that
do not solve  $\cD(\eta)H=0$. In the case that three $\eta$ values
coincide we get additional doubly-logarithmic solutions of
$\cD^3(\eta)H=0$ that solve neither $\cD(\eta)H=0$ nor
$\cD^2(\eta)H=0$.  A similar statement applies when $\Omega=0$ but
then the only critical case occurs when $\ell m^2= |\mu|$, in which
case we get a logarithmic solution of $\cD^2(\pm1)H=0$.  The catalog
of critical points for $\Omega\ne0$ is much
richer\footnote{The critical points of GMG have also been studied in \cite{Liu:2009pha}. Recently, properties of the logarithmic CFT dual to
GMG at the critical points have been investigated in
\cite{Grumiller:2010tj}.}:

\begin{enumerate}

\item $|\eta_+|=1$ but $|\eta_-|\ne 1$ and $\Omega^{-1}\ne0$. We find that $\eta_+ = \pm1$ when $\ell \mu \hat \sigma = \mp 1$ and then  $\eta_- = \mp 1/\left(1\pm \ell m^2/\mu\right)$. We get no propagating graviton mode from
$\cD(\eta_+)H=0$, just an  extra logarithmic solution to $\cD^2(\pm
1)H=0$. In this case there is therefore a {\it single} propagating
massive graviton of helicity $2\eta_-/|\eta_-|$. If $\Omega^{-1}
\rightarrow 0$, then $m^2\rightarrow\infty$, giving the TMG limit.
In this case $\eta_-=0$, and $|\ell\mu\sigma|=1$.

\item $|\eta_-|=1$ but $|\eta_+|\ne 1$ and $\Omega^{-1} \ne0$. In this case $\eta_- = \pm1$ when $\ell \mu \hat \sigma = \mp 1$ and then  $\eta_- = \mp 1/\left(1\pm \ell m^2/\mu\right)$. In this case there is  a {\it single}  propagating massive graviton of helicity
$2\eta_+/|\eta_+|$. If $\Omega^{-1} \rightarrow 0$, then
$m^2\rightarrow\infty$, giving the TMG limit. In this case
$\eta_+=0$, and $|\ell\mu\sigma|=1$.

\item $\eta_+ = -\eta_-$ and $|\eta_\pm|=1$. This case is realized by $\eta_-=-\eta_+=1$, which requires $|\mu|=\infty$, $m^2 \hat\sigma = 0$. This is the critical limit of NMG.  In this case (\ref{keyGMG}) degenerates to
\be\label{d22} \cD(1)^2\cD(-1)^2H=0\, . \ee Although this propagates
no gravitons, it does propagate two spin-$1$ modes
\cite{Bergshoeff:2010mf}. One way to see this is to rewrite the
fourth-order equation as the pair of second-order equations \be
\left(\ell^2\bar D^2 +2\right) H = U \, , \qquad \left(\ell^2\bar
D^2 +2\right)U=0\, , \ee where $U$ is another symmetric traceless
and `divergence-free' tensor.  Applying to the $U$-equation the same
reasoning that we used above for the `Einstein'  case, we deduce
that  $U_{\mu\nu} = 2\bar D_{(\mu}A_{\nu)}$ for some vector field
$A$ satisfying $\left(\ell^2\bar D^2-2\right)A_\mu =0$ and, since
$U$ is traceless,  $\bar D^\mu A_\mu =0$. In other words $H$
satisfies
\be\label{three2} \left(\ell^2\bar D^2 +2\right)H_{\mu\nu} = 2\bar
D_{(\mu}A_{\nu)}
\ee for a vector field such that \be
\left(\ell^2\bar D^2-2\right)A_\mu =0\, , \qquad \bar D^\mu A_\mu
=0\, . \ee These equations for $A$, which also follow from taking
the  divergence and trace of (\ref{three2}), are Proca equations in
the adS background. They are derivable from the Proca Lagrangian
density
\be
{\cal L}_{Proca} = -\ft14 F_{\mu\nu}F^{\mu\nu} -2 \ell^{-2} A^\mu A_\mu\ ,\qquad
F_{\mu\nu}= 2\partial_{[\mu}A_{\nu]}\ .
\ee
For singleton solutions the left-hand-side of \eqref{three2} vanishes, implying that $A$ is a Killing vector of the adS background.  In contrast,  the left-hand-side of \eqref{three2} is non-vanishing for the logarithmic solutions of \eq{d22} and hence these must be related to
non-trivial Proca modes. However, although a solution of the Proca equations furnishes a UIR of the adS$_3$  isometry group,
with $(E_0,s)= (2,1)$, a careful study of \eq{three2} shows that these solutions correspond to the first descendants of the logarithmic modes; see \cite{Giribet:2008bw} for a detailed description of precisely such a descendant mode.

\item $\eta_+=\eta_-$ but $|\eta_\pm|\ne1$. This is possible only when $\Omega>0$. In this case, $\eta_\pm = -2\mu/(m^2 \ell)$. We get one propagating massive graviton from the solution to $\cD( -2\mu/(m^2\ell))=0$ and an additional logarithmic solution of
$\cD^2( -2\mu/(m^2\ell))=0$.

\item  $\eta_+=\eta_-$ {\it and} $|\eta_\pm| =1$. This is possible only when $\Omega=1$, which requires $2\mu =\mp m^2 \ell$ according to whether $\eta_+=\eta_- =1$ or $\eta_+=\eta_- =-1$. In this case we have, apart from singletons, only logarithmic and doubly logarithmic modes from $\cD^3(1)H=0$ or $\cD^3(-1)H=0$.

\end{enumerate}

Observe that only cases (i)-(iv) are possible for $\Omega<0$, which may be required if ghosts are to be avoided, for the reason  given above.  In these four cases we have
\be
c_+ c_- =0\, .
\ee
In other words, at least one of the two central charges of the boundary CFT is zero. Unless both are zero,  ``critical'' can be interpreted as ``chiral''.  As pointed out in \cite{Liu:2009pha}, the ``chiral GMG''  models interpolate between chiral TMG, which is case (iii), and critical NMG, which is case (iv).

In the remainder of this paper, we shall explain how these results extend to the supergravity case. The ${\cal N}=1$ supersymmetric extensions of NMG and GMG have been found in recent work, along with all  possible maximally supersymmetric vacua \cite{Andringa:2009yc,Bergshoeff:2010mf}. These constructions are complicated by the fact that the off-shell graviton multiplet contains not just the metric but also an additional scalar field $S$, which is auxiliary in 3D `Einstein'  supergravity \cite{Howe:1977ut} and in super-TMG \cite{Deser:1982sw} but which propagates in generic curvature-squared models. Even when the kinetic term for $S$ is absent  it will still propagate a scalar mode in adS vacua unless the coefficients of the cubic equation of motion for $S$ are constants\footnote{This condition may be relaxed slightly but we refer to \cite{Bergshoeff:2010mf} for a discussion of this point.}, and this
condition defines  a ``super-GMG'' model for which the bosonic action is precisely of GMG-form after elimination of $S$. Remarkably, {\it all}  adS vacua of GMG correspond to supersymmetric vacua of super-GMG, so the spectrum in such vacua is determined by the GMG spectrum.   Of course,  the boundary CFT must now be a boundary SCFT, but it will still have the same central  charge \cite{Banados:1998pi}.  For these reasons, the results summarized above for GMG extend with no essential modifications to super-GMG. Unfortunately,
this makes it unlikely that supersymmetry can help to resolve the boundary/bulk unitarity ``clash''  explained above.

Although the super-GMG model is the most interesting special case of the generic curvature squared 3D supergravity model
constructed in \cite{Bergshoeff:2010mf},  the results for the generic model provide us with the opportunity to explore some issues in the context of a model with more parameters. The bosonic Lagrangian is
\be
\begin{split}
e^{-1} {\mathcal L}_{bos} &= -V(S) + f(S)R +\frac{1}{m^2}
\left(R^{\mu\nu}R_{\mu\nu} - \frac{3}{8}R^2\right) +
\frac{1}{8\tilde m^2} \left[R^2 - 16\left(\partial S\right)^2\right]
+ \frac{1}{\mu} e^{-1} {\cal L}_{LCS}\ , \label{genbos}
\end{split}
\ee
where ${\cal L}_{LCS}$ is the Lorentz-Chern-Simons term, and
\be
\begin{split}
V(S) &= -MS+2\sigma S^2 -\frac{1}{\check\mu} S^3 +\left(-\frac{3}{2m^2}+\frac{1}{\check m^2} +\frac{9}{2\tilde m^2}\right) S^4\ , \\
f(S) &= \sigma +\frac{1}{2\check\mu}S + \left(-\frac{1}{2m^2}+\frac{3}{10\check m^2} +\frac{3}{2\tilde m^2}\right) S^2\ .
\label{Vf}
\end{split}
\ee
The model has six  independent mass parameters $(M, m, \check m,
\tilde m, \check\mu,\mu)$, which can be traded for dimensionless
parameters using the gravitational coupling constant, and one
discrete parameter $\sigma=\pm 1,0$. Despite  the notation, $(m^2,
\tilde m^2,\check m^2)$ are allowed to take either sign.

The field equations following from \eq{genbos} are given in \cite{Bergshoeff:2010mf}.  The supersymmetric adS vacuum solutions of these equations have\footnote{We choose the sign of $S$ so as to agree with the sign choice made  in \cite{Bergshoeff:2010mf}.}
\be
S=\bar S = - \ell^{-1}\, , \qquad  M= - \ell^{-1}\left(4\sigma + \frac{2}{5\ell^2\check m^2} \right)\, .
\label{vac}
\ee
The central charges are given by \eqref{ccfun} with $\hat\sigma$ given by
\be\label{newsighat}
\hat\sigma = \sigma -\frac{1}{2\ell\check\mu}  +\frac{3}{10\ell^2\check m^2}\, .
\ee
The adS/CFT correspondence  suggests  that  the spectrum of propagating gravitons will be independent of $\tilde m$ and will depend  on the $\check\mu$ and $\check m$ only through the parameter $\hat\sigma$. We shall verify this.

To expand the field equations about a supersymmetric adS vacuum, we
write the metric as in (\ref{perturb}) and we write \be S= \bar S+
\ell^{-1}s \ee for dimensionless scalar perturbation  $s$.  To
present the results,  we shall again make use of the parameters
$\eta_\pm$ and $\Omega$  defined in  (\ref{Omega}) and
(\ref{etapm1})  in terms of $\hat\sigma$, although it should be
remembered that $\hat\sigma$  is now given by (\ref{newsighat}).  In
addition, it is useful to define the  dimensionless parameter \be a
= m^2 \left[ - \frac{2}{m^2} + \frac{6}{5\check m^2} +
\frac{6}{\tilde m^2} - \frac{\ell}{\check\mu} \right]\, . \ee This
agrees with the definition in \cite{Bergshoeff:2010mf} when
restricted to the models (with $\tilde m^2=\infty$) for which this
parameter was defined there. We will also need the following
(dimensionless) linear differential operators: \be L = \ell^2\bar
D^2 -3\ ,\qquad  \tilde L  = \frac{m^2}{\tilde m^2} L - \Omega \,  .
\label{LZJ} \ee Using these definitions, one finds that the $h$ and
$s$ equations are, respectively, \be
 \frac{1}{m^2}L\left(\tilde L  h + 3a s\right) =0\, , \qquad
\frac{1}{m^2} \left(\tilde L + 2a\right) s + \frac{a}{12 m^2} L h =0\, ,
\label{feqs2}
\ee
and that the linearized $H$-equation is
\be\label{feqs1}
 \cD(1)\cD(-1)\cD(\eta_+) \cD(\eta_-) H =\Omega^{-1}\, J\ , \qquad
\ee where \be J_{\mu\nu} = \frac{1}{3\ell^2} \left( \bar D_{\mu}
\bar D_{\nu} -\ft13 \bar g_{\mu\nu} \bar D^2 \right) \left[ \tilde L
h + 3as\right] \, . \ee The  integrability condition $\bar D^\mu
J_{\mu\nu}=0$ is satisfied as a consequence of the $h$-field
equation.

Observe that we recover the GMG equations that we have already
analyzed on setting  $\tilde m^2=|\check\mu|=\infty$ and $a=0$. In that case, we
argued that it was possible to set $h=0$, and hence $J=0$,  by using
a combination of the $h$-equation and a residual gauge
transformation. It is no longer so clear that this argument still
applies, but a simpler one is available, as long as $\Omega\ne0$. We
define the new symmetric traceless and `divergence-free'
perturbation \be \tilde H = H -m^2 \Omega^{-1} J \, . \ee Using the
relation $\varepsilon_{(\mu}{}^{\alpha\beta} {\bar D}_{|\alpha}
J_{\beta|\nu)} =0$, we may deduce that \be \cD(1)\cD(-1)\cD(\eta_+)
\cD(\eta_-) \tH_{\mu\nu}=0 \ . \label{key1} \ee The analysis of
critical points for $\Omega\ne0$ now proceeds {\it exactly} as for
the GMG case. For $\Omega=0$ the $H$-equation is given by
\be \left[\cD(\eta) {\cal D}(1){\cal D}(-1)\right]_\mu{}^\rho
\varepsilon_\rho{}^{\alpha \beta} \bar{D}_\alpha H_{\beta \nu} =
-\eta J_{\mu \nu}\, , \qquad \eta= - \frac{\mu}{\ell m^2}\, . \ee
Note that by acting on this equation with the operator
$\varepsilon_\lambda{}^{\tau\mu}\bar D_\tau$, we can arrive at the
integrability condition $\cD(\eta) {\cal D}(1){\cal D}(-1)(\ell^2
\bar{D}^2 + 3)H=0$, an equation that also follows from
(\ref{partmass}). To summarize, the critical point structure of the
generic curvature-squared supergravity model, expanded about a
supersymmetric adS vacuum,  is {\it identical}  to that  of the GMG
model, with all dependence on the extra parameters absorbed into the
effective EH coefficient $\hat\sigma$, as anticipated above.

The main difference of the generic case as compared to GMG is  the possibility of additional scalar modes arising from the $h$ and $s$ equations (\ref{feqs2}). Here we shall consider some special cases.

\begin{itemize}

\item $\tilde m^2=\infty$ {\it and} $a=0$; this case was called ``generalized GMG''  in \cite{Bergshoeff:2010mf}; it reduces to the
GMG case analysed above when $|\check\mu|=\infty$.   For this case, and assuming non-zero finite $m^2$,  the  equations (\ref{feqs2}) become
\be \Omega Lh =0\, , \qquad
\left(2a-\Omega\right) s=0
 \ee
 We see that $s=0$ (unless $\Omega=2a$ but then it is undetermined, implying  an `accidental'
gauge invariance that allows one to choose
$s=0$). Provided
$\Omega\ne0$ we have $Lh=0$ and residual gauge invariances allow us
to set $h=0$.  There are therefore no propagating scalars. One can show that the generalized super-NMG limit ($|\mu|=\infty$) is ghost-free for $\hat\sigma \le 0$ \cite{Bergshoeff:2010mf}.

\item $\tilde m^2=\infty$ but $a\ne0$. In this case the $S$-equation of motion is algebraic but it still propagates modes in adS, for reasons explained in \cite{Bergshoeff:2010mf}. Here we verify this for the generic $\Omega\ne0$ case. Under these conditions, the equations (\ref{feqs2}) become equivalent
to (for $m^2 \ne 0$ and finite) \be L \left(h- 3a\Omega^{-1}
s\right)=0 \, , \qquad L s =
\frac{4\Omega\left(\Omega-2a\right)}{a^2} s \ee As the addition of a
scalar to $h$ does not change its residual gauge transformation, the
first equation does not lead to propagating modes. The second one
takes the form (\ref{KGeq}) with $\varphi=s$ and $\mathcal{M}^2= 3 +
4\Omega\left(\Omega-2a\right)/a^2$. This equation therefore
propagates a scalar mode that is non-tachyonic.  However  there is no guarantee that
it  is not a ghost.

\item $a=0$ but $\tilde m^2 \ne \infty$. The equations (\ref{feqs2}) now reduce to
\be L \left(L- \frac{\Omega \tilde m^2}{m^2}\right)h =0\, , \qquad
\left(L- \frac{\Omega \tilde m^2}{m^2}\right)s =0\, . \ee The first
of these is a fourth-order equation for $h$ that is not easy to
analyse. The second equation takes the form (\ref{KGeq}) with
$\varphi=s$ and $\ell^2\mathcal{M}^2=3+\tilde{m}^2 \Omega/m^2$. We
therefore get a non-tachyonic scalar provided that $4\ge-\tilde{m}^2
\Omega/m^2$, although there is again no guarantee that it is not a
ghost.

\end{itemize}


\ack

EB wishes to thank the organizers for a stimulating and inspiring atmosphere. PKT is supported by an EPSRC Senior Fellowship. The research of ES is supported in part by NSF grants PHY-0555575 and PHY-0906222.  The work of O.H. is supported by the DFG--The German Science Foundation and in part by funds provided by the U.S. Department of Energy (DOE) under the cooperative research agreement DE-FG02-05ER41360.

\end{document}